**Controlled polymorphic competition – a path to tough and hard ceramics**

D.G. Sangiovanni,[1*] A. Kjellén,[1] F. Trybel,[1] L.J.S. Johnson,[2] M. Odén,[1]
F. Tasnádi,[1] I.A. Abrikosov[1]

[1] Department of Physics, Chemistry, and Biology (IFM), Linköping University, SE 58183 Linköping, Sweden
[2] Sandvik Coromant, 126 80 Stockholm, Sweden

From nanoscale devices including sensors, electronics, or biocompatible coatings to macroscale structural, automotive or aerospace components, fundamental understanding of plasticity and fracture can guide the realization of materials that ensure safe and durable performance. Identifying the role of atomic-scale plasticity is crucial, especially for applications relying on brittle ceramics. Here, stress-intensity-controlled atomistic simulations of fracture in cubic $Ti_{1-x}Al_xN$ model systems demonstrate how *Å-scale* plasticity — manifested as lattice distortions, phase transformation, nucleation and emission of dislocations — substantially affects the *macroscale* fracture toughness ($K_{Ic}$) and fracture strength ($\sigma_f$) of brittle ceramics. The extent of plastic deformation in $Ti_{1-x}Al_xN$ increases monotonically with the Al content (x), due to a corresponding decrease in cubic → hexagonal polymorph transition energies and unstable stacking fault energies. Overall, plasticity positively affects the mechanical properties, resulting in optimal combinations of strength and toughness for x≈0.6. However, for x exceeding ~0.7, the benefits of plasticity diminish. The initial rise followed by a decline in $K_{Ic}(x)$ and $\sigma_f(x)$ is explained based on the interplay between phase transformation, shear-induced faulting, and tensile cleavage on the easiest fracture plane. The results highlight the impact of atomic-scale plasticity on observable properties and point to strategies for toughening ceramics through control of polymorph competition.

***davide.sangiovanni@liu.se**

## 1. Introduction

Ceramics are indispensable for technologies that demand a combination of high hardness, heat tolerance, and good chemical inertness. Unfortunately, they are often characterized by a tendency to fracture before yielding. Brittleness is a problem of major concern, especially in safety-critical applications. In contrast to most ceramics, zirconia ($ZrO_2$) and $ZrO_2$-based alloys [1] display exceptional resistance to fracture, with measured fracture toughness values ($K_{Ic}$) up to 20 MPa √m [2-4]. Such unusual mechanical behavior stems from stress-activated phase transformation ahead of growing cracks, which effectively prevents crack propagation [5-7]. However, the relatively low hardness (H ≈10 GPa [3, 8, 9]) and rapid softening with temperature [8-10] makes $ZrO_2$-based alloys unsuited for many ceramics or refractory applications.

The tetragonal-to-monoclinic transition responsible for transformation-toughening in $ZrO_2$-based ceramics is accompanied by an anisotropic volumetric strain and shear deformation [11]. At room temperature, the transformation is driven by stress, with transition enthalpies of the order of few tens of meV/atom [12, 13]. These characteristics suggest that control of the thermodynamic competition between polymorph phases can allow designing refractory ceramics with high hardness combined to improved toughness. The design concept is verified here for transition-metal nitride (TMN) alloys.

TMN ceramics are widely applied as protective thin-film coatings on, e.g., cutting tools, engine components, and medical instruments due their high hardness (H ≈ 20 – 45 GPa [14-16]), fracture strength ($\sigma_f$ ≈ 2 – 6 GPa [17-20]), and good chemical stability. Similar to other refractory materials, they are characterized by brittleness and low toughness ($K_{Ic}$ ≈ 1 – 5 MPa √m [16]). For our analysis, we consider rocksalt-structured (B1) pseudobinary $Ti_{1-x}Al_xN$ solid solutions as representative model systems because Al substitutions enable tuning of the competition between different polymorphs within a relatively narrow energy range [21, 22].[f1] Furthermore, experiments have shown that AlN-containing alloys and superlattices exhibit improved fracture resistance [28-30], phenomenologically attributed to stress-activated transformation plasticity in AlN-rich domains [28-30].

Processes that govern plasticity and fracture are often beyond the spatial and/or time resolution of electron microscopy. Experimental identification of nanoscale plasticity becomes even more complex for thin-film coatings, such as $Ti_{1-x}Al_xN$ [31, 32], due to small sample sizes and substrate effects. We employ atomistic simulations implemented with stress-intensity-controlled deformation of defective crystal models to characterize the fracture strength and fracture toughness of $Ti_{1-x}Al_xN$ ceramics. The availability of experimental fracture properties – measured by $Ti_{1-x}Al_xN$ microcantilever bending and micropillar splitting over a wide compositional range – allows supporting theoretically-determined trends and hence ascribing the observed toughening and strengthening effects to plasticity at the atomic scale.

## 2. Computational methods

Density-functional theory (DFT) calculations and *ab initio* molecular dynamics (AIMD) simulations are carried out using the VASP code implemented with projector augmented-wave pseudopotentials [33, 34]. In DFT calculations, the energy cutoff of planewaves is set to 400 eV. The thickness of k-point grids used for integration of the reciprocal space varies depending on the property under investigation (details are given in the **Supplemental Information** (**SI**)). AIMD simulations employ Γ-point sampling of the Brillouin zone and 300 eV cutoff energies. In both DFT and AIMD, the electronic exchange and correlation energies are treated with the approximation of Perdew-Burke-Ernzerhof [35].

[f1] Note that B1 (Ti,Al)N solid solutions are thermodynamically inclined to spinodal decompose into cubic AlN-rich/TiN-rich coherent domains [23]. However, spinodal decomposition is kinetically-prevented by very low atomic diffusivities in nitrides [24–27].

For calculations of surface energies at 0 K, we also employ the local density approximation (LDA). Classical molecular dynamics (CMD) and molecular statics (MS) simulations are performed with LAMMPS [36] describing Ti-Al-N interactions based on the second-neighbor modified-embedded atom method (MEAM) [37], as parameterized for Ti-Al-N systems in Ref. [38]. Videos and figures showing atomistic structures are created using VMD [39]. Description of the videos can be found in **Section S4 of the SI**.

### 2.1. Tensile deformation of notched lattice models

The computational approach used in CMD and AIMD simulations of tensile-testing of small notched lattice models have been detailed in our previous work [40]. The structures contain 1100 atoms with a supercell area of ≈13 $nm^2$ parallel to the [001] strain direction (see **Figure 1**). The simulation box is elongated up to fracture, incrementing strain by 2% every 1.5 ps. At each strain step, the structures are equilibrated via NVT sampling (300 K) with a timestep of 1 fs.

### 2.2. K-controlled loading of cracked plate models

The method used to carry out stress-intensity ($K_I$) controlled simulations of fracture of a cracked plate lattice model (schematically illustrated in **Figure S1 of the SI**) has been implemented in our previous study on TiN [40]. The plate models contain an atomically sharp crack on the (001) or on the (111) plane (with crack front along the [010] and $[1\bar{1}0]$ crystal axis, respectively) and are periodic along the crack front direction. The (001) surface is the one of lowest formation energy in B1-structured ceramics [41] and is thus expected to be the easiest path for transgranular fracture. Although the (111) surface of B1-structured $Ti_{1-x}Al_xN$ has a much higher formation energy than the (001) [42, 43], the texture of polycrystalline $Ti_{1-x}Al_xN$ PVD coatings is dominated by (001)- and (111)-faceted grains [44-47]. Therefore, fracture during experimental mechanical testing could initiate at nanovoids or flaws with local (111) or

(001) surface orientation. In $Ti_{1-x}Al_xN(111)$ plate models, atomically-sharp (111) cracks are created by removing half Ti monolayer, which results in two N-terminated surfaces on opposite sides of the crack plane. The choice is motivated by the fact that N-terminated TiN(111) surfaces are energetically more stable than Ti-terminated (111) surfaces [48]. The stress-intensity factor is incremented stepwise ($\Delta K_I$ = 0.02 MPa√m) by controlling the displacement of atoms within the frame region of the plate model [40, 49] (see also **Figure S1**). Crack-healing is prevented by screening the interactions of atoms on opposite sides of the crack plane.

$K_I$-controlled simulations reveal that B1 $Ti_{1-x}Al_xN$ alloys with Al concentrations x > 0.7 undergo phase transformation by nucleation and growth of the $B_k$ hexagonal structure around the crack front. The stress-intensity required to activate such transformation decreases rapidly for Al metal percentage increasing above 70%. This observation requires separate characterization of the mechanical properties of the hexagonal phase. Thus, we compute the equilibrium structure and elastic tensor of configurationally-disordered $B_k$ $Ti_{0.25}Al_{0.75}N$, $Ti_{0.15}Al_{0.85}N$, and $Ti_{0.05}Al_{0.95}N$ alloys at 0 K. The calculated properties serve as input to $K_I$-controlled simulations of fracture of the hexagonal alloys.

In atomistic modelling of fracture, the critical stress intensity $K_{Ic}$ corresponds to the onset of unstable crack growth. For an ordered brittle crystal, such as TiN, the atomically-sharp crack extends on a straight path [40] at a stress intensity $K_{Ic} \approx (1+\epsilon)\cdot K_{Ic}^G$ , where $K_{Ic}^G$ is the thermodynamic Griffith value of mode-I toughness and the scalar $\epsilon$ (≈0.2) arises from lattice trapping of brittle cracks [50-52]. At variance with Ref. [40], here we have developed an algorithm that follows the movement of the crack tip. This allows us to determine conditions of unstable crack growth in $Ti_{1-x}Al_xN$. In these alloys, the formation of plastic zones complicates the analysis of $\sigma_f$ and $K_{Ic}$ fracture properties. At each $K_I$ step, the position of the crack tip is (re)located considering the maximum atom-resolved stress and geometric descriptors that control changes in bond angles and bond lengths proximate to the identified position of maximum stress. This ensures an accurate definition of $K_I$ during crack extension [49]. **Videos #1 and #5, #6, and #7** illustrate how the algorithm traces the crack tip movement in TiN(001) and $Ti_{0.4}Al_{0.6}N$(001).

### 2.3. Determination of $K_{Ic}^{\infty}$ and $\sigma_f^{\infty}$

$K_I$-controlled MS simulations provide tensile stress $\sigma_{zz}$ vs stress-intensity $K_I$ data (**Figure S10 and S11**) for each simulated plate model of area A. The $\sigma_{zz}(K_I)$ curve exhibits linear dependence on $K_I$ up to stress intensity values near the onset of plastic deformation or fracture. We analyze $\sigma_{zz}$ vs. $K_I$ data for stress intensities in the range $0 \leq K_I \leq K_I^{5\%}+\Delta K$ (with $\Delta K \ll K_I^{5\%}$). The threshold value $K_I^{5\%}$ corresponds to a lateral advancement of the crack tip ($\Delta_{tip}$), or extension of the secondary $B_k$ phase, equal to 5% of the supercell width. The criterion is based on the observation that all investigated alloys exhibit either unstable crack extension or rapid phase transformation at $K_I \approx K_I^{5\%}$. Then, we locate the maximum stress $\sigma_{zz}^{max}(\widetilde{K}_I)$ value within the stress-intensity interval $0 < \widetilde{K}_I \leq K_I^{5\%}$. This allows us to define the fracture strength $\sigma_f \equiv \sigma_{zz}^{max}(\widetilde{K}_I)$ and fracture toughness $K_{Ic} \equiv \widetilde{K}_I$ for each plate area *A*. Visualization of the simulations, supported by analyses of $\sigma_{zz}$ and $\Delta_{tip}$ trends as a function of $K_I$, confirms that the definitions of $\sigma_f$ and $K_{Ic}$ are physically meaningful.

Atomistic predictions of fracture properties are spuriously affected by limited supercell sizes and boundary conditions [53]. Here, we circumvent these problems by explicitly considering the dependence of fracture properties on the size of supercell models. As expected, the calculated fracture properties vary smoothly with the area of the cracked plate. Accordingly, the fracture toughness and fracture strength reach asymptotic values $K_{Ic}^{\infty}$ and $\sigma_f^{\infty}$ for $A \rightarrow \infty$. Spurious effects caused by boundary conditions are also expected to vanish at the limit of infinite sizes. Hence, $K_{Ic}^{\infty}$ and $\sigma_f^{\infty}$ are extrapolated by using constitutive scaling laws that reproduce trends in $K_{Ic}$ and $\sigma_f$ as a function of *A* (see also **Section S1 of the SI**). Importantly, the plasticity and fracture mechanisms identified in our simulations remain qualitatively unchanged when changing the area of the cracked plate models. This strengthens our confidence that the mechanical behavior observed in atomistic models accurately represents the stress response of real alloys. Consequently, data collected from finite sizes can be reliably used to extrapolate macroscale properties.

The values $K_{Ic}^{\infty}$ obtained by atomistic simulations of fracture are compared to the thermodynamic value $K_{Ic}^{G}$ based on Griffith's theory of fracture. Griffith's formulation assumes a perfectly homogeneous elastic body containing a crack. Hence, the value of $K_{Ic}^{G}$ for the cracked plate problem is simply obtained from the cleavage energy and the Stroh energy tensor $\Lambda$, which can be calculated from the elastic constants tensor $C_{ij}$ [40, 49] (more information in **Section S2 of the SI**). The elastic constants predicted by MS for B1 $Ti_{1-x}Al_xN$ have been taken from Ref. [38]. In this work, we also compute the elastic constants of hexagonal $B_k$ $Ti_{0.25}Al_{0.75}N$, $Ti_{0.15}Al_{0.85}N$, and $Ti_{0.05}Al_{0.95}N$. The unrelaxed surface energies of $Ti_{1-x}Al_xN(001)$ are computed by MS and DFT using supercells with large areas (to minimize degree of short-range lattice ordering) and thickness of 6 layers. Calculations of surface energies are done as previously (see, e.g. [54]). The results of surface energies calculated by MS and DFT are summarized in **Table S1 of the SI**.

### 2.4. Transformation energies and degree of plastic deformation

The percentage of cubic and transformed (non-cubic) structures in mode-I deformed $Ti_{1-x}Al_xN$ cracked plate models is determined using the common neighbor analysis (cutoff of 3.15 Å) as implemented in OVITO [55]. The energies of transformation from cubic B1 to hexagonal $B_k$ are determined as a function of the Al content along a transformation path schematically illustrated in **Figure 1c**. The calculations at a given Al content are repeated with several different cation arrangements. The B1→$B_k$ transformation energy landscapes are depicted in **Figure S6 of the SI**. Results obtained with the classical potential are supported by DFT calculations, which qualitatively show that the $\Delta E_{B1\rightarrow Bk}$ activation energy decreases for increasing Al content (**Figure S7**).

## 3. Results and discussion

In atomistic simulations of fracture, an accurate evaluation of mechanical properties can be assured by controlling the stress intensity factor ($K_I$) at the crack tip of a defective lattice while also verifying that inelastic processes are confined to a relatively small region centered at the tip [49]. Mode-I (crack-opening) and mode-II (shearing) loading based on anisotropic elasticity has been implemented in our previous study on the competition between dislocation emission and crack propagation in $TiN_x$ [40]. Here, we investigate B1 $Ti_{1-x}Al_xN$ ($0 \leq x \leq 0.95$) crystal models that contain an atomically sharp crack on the (001) or on the (111) plane. Below, the defective lattices with different crack geometry are indicated as $Ti_{1-x}Al_xN(001)$ or $Ti_{1-x}Al_xN(111)$. We note that atomically-sharp cracks have been experimentally observed in diverse material systems [56-59]. **Section S1** of the **SI** details supercell models and methods used in atomistic simulations of fracture.

### 3.1. Validation of the potential against *ab initio* results for notched lattice models

The classical MEAM potential – as parameterized in Ref. [38] – has been proven to reproduce experimental and *ab initio* properties of defect-free and defective TiN and Ti-Al-N alloys, including their different polymorph structures [40, 60-62]. The interested reader may consult **Section S2 of the SI** for detailed information on properties and phenomena that have been tested and validated. To verify that the MEAM classical potential can realistically describe plastic deformation and fracture in $Ti_{1-x}Al_xN$ alloys (prior to carrying out computationally-intensive $K_I$-controlled simulations), we compare the results of classical molecular dynamics and *ab initio* molecular dynamics obtained for small, notched supercell models subjected to ⟨001⟩ tensile elongation. Besides serving for validation of the MEAM model, the simulations allow us to rapidly probe the tendency of B1 $Ti_{1-x}Al_xN$ alloys to deform plastically.

CMD and AIMD results of deformed notched $Ti_{0.5}Al_{0.5}N$ and $Ti_{0.25}Al_{0.75}N$ models are presented in **Figure 1**. Our previous simulations showed that notched TiN fractures in a brittle manner at ~30% tensile strain of the simulation box (figure 5 in [40]). At the same elongation, B1 $Ti_{0.5}Al_{0.5}N$ and $Ti_{0.25}Al_{0.75}N$ undergo lattice transformation around the notches (**Figure 1**). The observed structural changes preserve the integrity of the bonding network, thus indicating that $Ti_{0.5}Al_{0.5}N$ and $Ti_{0.25}Al_{0.75}N$ are qualitatively more plastic and resistant to fracture than TiN. The lattice distortions seen in $Ti_{0.5}Al_{0.5}N$, resemble the initial stage of cubic B1→hexagonal $B_k$ transformation (see orange circles and dotted lines in **Figure 1a,c**). For the case of notched $Ti_{0.25}Al_{0.75}N$, a strain of 50% leads to the formation of $B_k$-structured domains (**Figure 1b,c**). The $B_k$ phase (hexagonal BN prototype) is a honeycomb-patterned variant of the hexagonal wurtzite (B4) structure in which metal and N atoms of adjacent (0001) layers shift onto the same lattice plane (**Figure S8**). Accordingly, during B1→$B_k$ phase transformation, the atomic coordination decreases from six- to five-fold.

The formation of $B_k$-structured domains in strained notched $Ti_{1-x}Al_xN$ models is surprising, as the $B_k$ phase has (to the best of our knowledge) not been previously detected in $Ti_{1-x}Al_xN$ samples. Reports in the literature show that $Ti_{1-x}Al_xN$ films with compositions $x \gtrsim 0.7$ often exhibit a dual phase (B1 + B4) or single phase (B4) structure [63].[f2] Oppositely, $Ti_{1-x}Al_xN$ films with Al contents $x<0.7$ are typically single-phase B1 [31]. The experimental findings are consistent with the results of DFT calculations indicating that $Ti_{1-x}Al_xN$ solid solutions energetically favor the B4 (or B1) structure for Al contents greater (or lower) than ~0.7 [21]. DFT investigations also demonstrated that the B4 phase is mechanically unstable for

[f2] $Ti_{1-x}Al_xN$ films synthesized by physical vapor deposition exhibit single-phase B1 structure for x up to ~0.7 [31]. However, results of experiments based on chemical vapor deposition indicated that single-phase B1 $Ti_{1-x}Al_xN$ solid solutions can be realized with Al contents as high as ~0.9 [32,64].

alloys with x≲0.6 [63], and relax into the $B_k$ during structural optimization [65]. Thus, previous DFT results [21, 63, 65] clarify why $B_k$ domains can form in B1 $Ti_{0.5}Al_{0.5}N$ under loading (**Figure 1a**). The appearance of $B_k$ (instead of B4) environments in strained $Ti_{0.25}Al_{0.75}N$ (**Figure 1b**) is even less expected than in $Ti_{0.5}Al_{0.5}N$. Our simulations suggest that the $B_k$ phase is stabilized through tensile elongation. The hypothesis is confirmed by complementary DFT and MS calculations showing that [1120]-elongated B4 $Ti_{0.25}Al_{0.75}N$ transforms into $B_k$-structured $Ti_{0.25}Al_{0.75}N$ (**Figure S9**). The results emphasize the relevance of $B_k$ domains in plastically deformed regions of Al-rich $Ti_{1-x}Al_xN$ ceramics.

The agreement between AIMD and CMD results of stress-activated transformation in notched lattices, together with the extensive validation presented in **Section S1 of the SI**, lends confidence that the MEAM force field can realistically describe plasticity and fracture processes in $Ti_{1-x}Al_xN$.

### 3.2. Influence of polymorphic transformation on properties of B1 $Ti_{1-x}Al_xN(001)$

Previous experimental studies observed an enhanced resistance to fracture in ceramic systems that contain AlN. It has been suggested that the improvement in mechanical properties originates from transformation plasticity in AlN-rich domains [28-30]. However, experimental evidence of the initiation of phase transition at crack tips is lacking. This motivates the use of atomistic simulations, which locate and elucidate plasticity mechanisms activated with loading.

$K_I$-controlled simulations of fracture are carried out for cracked plate supercell models containing up to 1.3 million atoms; a size deemed sufficient to reliably extrapolate macroscale properties (details in **Section S1 of the SI**). Simulations performed for $Ti_{1-x}Al_xN$ are repeated with different stochastic cation arrangements to gain reasonable statistical confidence on $K_{Ic}(A)$ and $\sigma_f(A)$ values for each plate area ($A$) considered here. The constitutive scaling laws identified in this work reproduce $K_{Ic}(A)$ and $\sigma_f(A)$ vs. $A$ asymptotic trends, and allow us to extrapolate the material fracture toughness $K_{Ic}^{\infty}$ and fracture strength $\sigma_f^{\infty}$ (intended as crack-initiation toughness and strength) at the macroscale limit. Such *macroscale* properties unambiguously characterize the mechanical behavior of B1 $Ti_{1-x}Al_xN$ with well-defined crack geometries. The properties measured for actual $Ti_{1-x}Al_xN$ samples are, instead, affected by mechanical-testing parameters and the structure, properties, and density of extended crystallographic defects.

The results of $K_I$-controlled deformation show that the mechanical properties of B1 $Ti_{1-x}Al_xN(001)$ and $Ti_{1-x}Al_xN(111)$ cracked plates vary with x in a qualitatively similar manner. We choose to focus on description of the properties of $Ti_{1-x}Al_xN(001)$ systems, as the {001} surfaces are often the most stable in B1 structured ceramics [41, 42, 66] and thus expected to be the easiest fracture planes. Noteworthy differences between the behaviors of cracks on (001) and (111) planes, and associated plastic deformation mechanisms, are described in **Section 3.3**.

**Figure 2a and 2c** report the values of $K_{Ic}^{\infty}$ and $\sigma_f^{\infty}$ obtained for B1 $Ti_{1-x}Al_xN(001)$ alloys. All simulations carried out for B1 TiN(001) evidence crack extension along the (001) plane (**Video #1**). The mechanical properties of TiN(001) extrapolated to the macroscale limit are $K_{Ic}^{\infty} \approx 1.8$ MPa $\sqrt{m}$ and $\sigma_f^{\infty} \approx 2.4$ GPa. As anticipated by results obtained for the small, notched lattice models in **Figure 1**, Al substitutions improve the resistance to fracture of $Ti_{1-x}Al_xN$ alloys. **Figure 2a,c** show a monotonic rise in the $K_{Ic}^{\infty}$ and $\sigma_f^{\infty}$ values for an aluminum content x increasing from 0 to 0.6. Accordingly, B1 $Ti_{0.4}Al_{0.6}N(001)$ shows the best combination of toughness (3.2 MPa $\sqrt{m}$) and strength (4.3 GPa). As clarified below, the enhancement in fracture resistance is due to lattice distortions localized around the crack tip and extrinsic toughening mechanisms. A further increase in aluminum concentration becomes less advantageous for the mechanical properties of $Ti_{1-x}Al_xN$ alloys. The effect is indicated by reductions in $K_{Ic}^{\infty}$ and $\sigma_f^{\infty}$ for x greater than 0.7 (**Figure 2a,c**). The deterioration in mechanical properties in Al-rich B1 $Ti_{1-x}Al_xN$ is caused by premature B1→$B_k$ phase transformation, activated under loading.

We should note that the formation of a secondary phase in the cubic host lattice introduces an uncertainty on $K_{Ic}$ and $\sigma_f$ values predicted for B1-structured $Ti_{1-x}Al_xN$ alloys with x>0.7. The definition of $K_I$ becomes progressively less accurate as the hexagonal phase grows within the B1 lattice.[f3] Nevertheless, the *true* fracture toughness and strength of B1 $Ti_{1-x}Al_xN$ with x > 0.7 are expected to be close to the values of **Figure 2a,c**. Our claim is motivated by the fact that the $K_I$-controlled simulations are terminated when the fractional advancement of the crack front ($\Delta_{tip}$), or extension of the secondary phase, has reached 5% of the plate model size (details in **Section 2.3**). Through separate analyses (not shown), we verify that setting a $\Delta_{tip}$ parameter below 5% does not alter the trends in mechanical properties as a function of the Al content. Additionally, the $K_{Ic}^{\infty}$ and $\sigma_f^{\infty}$ values of B1-structured alloys calculated for $0 < \Delta_{tip} < 5\%$ remain close to the confidence ranges shown in **Figure 2a,c**. An accurate determination of the properties of dual-phase $Ti_{1-x}Al_xN$ is beyond the scope of this work. However, complementary K-controlled simulations demonstrate that single-phase $B_k$-structured $Ti_{0.25}Al_{0.75}N$ and $Ti_{0.15}Al_{0.85}N$ are considerably less resistant to fracture than the corresponding cubic crystals (see **Figure 2a,c**, **Figure S11**, and **Videos #2 and #3**). Thus, the relatively high fracture toughness of B1 $Ti_{0.25}Al_{0.75}N(001)$ and $Ti_{0.15}Al_{0.85}N(001)$ are partially compromised by local transformations into the weaker $B_k$ phase.

The impact of atomic-scale transformation on the material's resistance to fracture can be assessed by comparing $K_{Ic}^{\infty}$ results of atomistic simulations (**Figure 2a**) with Griffith's $K_{Ic}^{G}$ values (**Figure 2b**). The latter are predicted from linear elastic fracture mechanics (LEFM). In LEFM, a solid is assumed to maintain a linear elastic response up to fracture (see **Section S2 in the SI**). Therefore, while a ratio $K_{Ic}^{\infty}/K_{Ic}^{G} \approx 1$ would correspond to perfectly brittle fracture, a $K_{Ic}^{\infty}/K_{Ic}^{G} \gg 1$ indicates that inelastic effects greatly contribute to the toughness. According

[f3] The $K_I$ is controlled through atomic displacements that are, in turn, calculated from the elastic response of the original untransformed cubic phase.

to LEFM, the fracture toughness of B1 $Ti_{1-x}Al_xN(001)$ would (misleadingly) decrease with increasing x (**Figure 2b**). The trends in $K_{Ic}^G$ vs. x obtained by MEAM are confirmed by DFT calculations (**Figure 2b**). On the contrary, $K_{Ic}^\infty$ results in **Figure 2a** demonstrate that Al substitutions have positive effects on the alloys' resistance to fracture, at least for x up to ≈0.6. The ratio $K_{Ic}^\infty/K_{Ic}^G$ (not plotted) increases monotonically from 1.2±0.1 (in brittle B1 TiN(001)) to 2.5±0.2 for B1 $Ti_{0.4}Al_{0.6}N(001)$, which is the toughest and strongest of the alloys tested in this work. The deviations between $K_{Ic}^\infty$ and $K_{Ic}^G$ trends qualitatively demonstrate the significant contribution of atomic-scale plasticity on the fracture toughness of the alloy.

Our simulations show that the extent of plastic deformation in B1 $Ti_{1-x}Al_xN(001)$ subjected to mode-I loading increases monotonically with the Al concentration. The effect is evidenced by structural analyses of $Ti_{1-x}Al_xN(001)$ cracked plate models at a stress intensity factor $K_I$ = 2 MPa √m (**Figure 3b-g**). The results obtained for notched models in **Figure 1** suggest that plasticity in mode-I loaded B1 $Ti_{1-x}Al_xN(001)$ lattices is dominated by the formation of $B_k$-like domains. The tension-driven B1→$B_k$ transformation path is represented in **Figure 1c**. Accordingly, the degree of plasticity should be inversely related to the polymorphic transition energy ($\Delta E_{B1\rightarrow Bk}$). The correlation between the degree of plasticity and $\Delta E_{B1\rightarrow B_k}^{-1}$ is shown in **Figure 3a**, where both quantities increase monotonically with the Al concentration x.

Irrespective of their size, TiN crystal models containing an atomically-sharp (001) crack undergo cleavage fracture when subjected to mode-I loading (**Figure 3b,b'**). However, as shown by results in **Figure 2a,c**, both strength and toughness can be enhanced by wisely tuning the metal composition. While a small addition of Al (x ≲ 0.25) in $Ti_{1-x}Al_xN(001)$ leads to relatively small improvements in fracture resistance (**Figure 3c,c'**), an Al content up to 0.6 induces progressively more pronounced structural distortions at the crack tip (**Figure 3d,d'**), which retards fracture by locally redistributing and/or dissipating stress. The enhanced resistance to crack propagation for x increasing from 0 to 0.6 is demonstrated by a slower crack-tip advancement $\Delta_{tip}$ as a function of $K_I$ (**Figure S10b**).

The mechanism for Al-mediated toughening and strengthening is detailed in **Figure 4**, where B1 $Ti_{0.4}Al_{0.6}N(001)$ is taken as representative case. **Figure 4a** and its magnification shows puckering of atomic layers around the crack tip. The corrugated portion of the $Ti_{0.4}Al_{0.6}N(001)$ lattice resembles $B_k$-like environments encountered along the B1 → $B_k$ transformation path. At the same time, the alloy undergoes amorphization at the crack tip (**Figure 4a**). A further increase in load does not lead to a complete transformation but rather initiates fracture. Nevertheless, the crack propagates relatively slowly in $Ti_{0.4}Al_{0.6}N(001)$ thanks to the formation of ligaments connecting opposite surfaces behind the crack tip (**Figure 4b** and **4c**). The dark blue color indicate that these "atomic bridges" carry substantial tensile load (**Figure 4b** and **4c**) and are therefore responsible for extrinsic toughening (crack-growth toughness [67]) of $Ti_{0.4}Al_{0.6}N(001)$. A similar phenomenon has been observed experimentally and proposed as toughening effect in other nitride ceramics [68].

Although the calculated toughness and strength of $Ti_{1-x}Al_xN(001)$ initially rise with x (**Figure 2a,c**), the white-gray color gradient in **Figure 3a** marks the compositional limit where the effects of plasticity on mechanical properties start becoming less beneficial. Indeed, **Figure 2a,c** evidences a decline in $K_{Ic}^{\infty}$ and $\sigma_f^{\infty}$ for x $\gtrsim$ 0.7. The reduction in strength and toughness originates from early mechanical yielding caused, in turn, by premature nucleation and growth of the weaker hexagonal $B_k$ polymorph. The phenomenon is exemplified for the case of $Ti_{0.15}Al_{0.85}N(001)$ in **Figure 5**. When subjected to mode-I loading, the B1 structure starts transforming into $B_k$ at a relatively low stress intensity factor ($K_I \approx 1.3$ MPa $\sqrt{m}$). We thus infer that an early onset of phase transformation (especially if extended) is detrimental for the alloy's strength and toughness. The scenario would qualitatively change if the newly formed phase were inherently stronger and tougher than the parent phase. This is the case for $Ti_{0.05}Al_{0.95}N(001)$, where the hexagonal structure exhibits better properties than the cubic one (**Figure 2a,c**). In absolute terms, however, $Ti_{0.05}Al_{0.95}N(001)$ alloys are predicted to possess relatively low toughness and strength (**Figure 2a,c**).

The trends in mechanical properties shown in **Figure 2a,c** can be rationalized from the competition between the polymorphic transition energy $\Delta E_{B1\rightarrow Bk}$ and the energy ($\gamma_f$) required to cleave defect-free B1 $Ti_{1-x}Al_xN(001)$ crystals during [001] uniaxial elongation. The quantities $\Delta E_{B1\rightarrow Bk}$ and $\gamma_f$ – readily obtained by static calculations of 3D periodic $Ti_{1-x}Al_xN$ supercells – are here identified as useful descriptors for the complex mechanical behavior of the alloy during mode-I loading. Hence, $Ti_{1-x}Al_xN$ lattice models with low, intermediate, and high Al contents are considered as representative cases to probe the effect of composition on the energetic tendency for brittle fracture or phase transformation. We compare energy trends computed during sequential [001]-elongation and stepwise deformation along the $B1\rightarrow B_k$ transformation path of **Figure 1c**. Details on the approach are given in **Section S3 of the SI**.

Considering the case of TiN – the most brittle among all systems studied here – **Figure 6a** shows that uniform uniaxial strain up to (001) cleavage is energetically favored over lattice transformation throughout the deformation process. The result is consistent with the observation that TiN(001) preserves octahedral B1-like coordination up to fracture during mode-I loading simulations (**Figure 3b,b' and Video #1**). Expressed in more general terms, for TiN(001) (and $Ti_{1-x}Al_xN(001)$ with relatively low Al content ($x\lesssim0.3$)), the energy required to activate transformation considerably exceeds the (001) cleavage energy $\gamma_f$ (**Figure 6d**). The large gap between $\gamma_f$ and $\Delta E_{B1\rightarrow Bk}$ aids our understanding for why these alloys exhibit rapid crack propagation during $K_I$-controlled simulations. Aluminum-rich ($x\gtrsim0.7$) $Ti_{1-x}Al_xN$ solid solutions display a diametrically opposite behavior. The energy accumulated by straining along the $B1\rightarrow B_k$ transformation path is constantly below the uniaxial elongation energy, as exemplified by results of $Ti_{0.15}Al_{0.85}N$ in **Figure 6c**. The fact that $\Delta E_{B1\rightarrow Bk} < \gamma_f$ for $x\gtrsim0.7$ (**Figure 6d**) may explain the tendency of $Ti_{1-x}Al_xN(001)$ alloys with high Al content to phase-transform before fracturing (see, e.g., **Figure 5**). Transformation plasticity would be obviously desirable if the newly formed phase were stronger than the original crystal structure. This is however not the case for $Ti_{0.25}Al_{0.75}N(001)$ and $Ti_{0.15}Al_{0.85}N(001)$, as the hexagonal lattice is weaker than the cubic one (**Figure 2a,c**). The crack front of alloys with intermediate Al concentrations undergoes a relatively slow and localized modification of the bonding network during $K_I$-controlled loading (see results of $Ti_{0.4}Al_{0.6}N(001)$ in **Figure 4a**). For Al concentrations $x \approx 0.6$, the values of tensile cleavage fracture $\gamma_f$ are slightly lower than $\Delta E_{B1 \rightarrow Bk}$ (**Figure 6b,d**). This suggests that alloys with intermediate Al contents can accommodate higher amounts of mechanical energy by lattice distortions before fracturing, which has an optimal effect on strength and toughness (**Figure 2a,c**).

### 3.3. Effects of stacking-fault formation on fracture resistance of B1 $Ti_{1-x}Al_xN(111)$

The effect of polymorphic competition on the mechanical properties of $Ti_{1-x}Al_xN$ ceramics is analyzed further by investigating fracture mechanisms in alloy models containing (111) cracks. MS simulations reveal that the mechanical response of $Ti_{1-x}Al_xN(111)$ alloys changes – from brittle to progressively more plastic – with increasing Al concentration, consistent with observations made for $Ti_{1-x}Al_xN(001)$. The trends in $K_{Ic}^{\infty}$ and $\sigma_f^{\infty}$ calculated for $Ti_{1-x}Al_xN(111)$ as a function of x are alike those of $Ti_{1-x}Al_xN(001)$, as indicated by black squares and purple triangles in **Figure 7**. Maximum toughness and strength of $Ti_{1-x}Al_xN(111)$ are recorded for Al concentrations around 0.6. It is, however, worth noting that the fracture toughness values calculated for $Ti_{1-x}Al_xN(001)$ are systematically lower (between ~10 and 30%) than those obtained for $Ti_{1-x}Al_xN(111)$ (**Figure 7a**). In contrast, the theoretical fracture strength of B1 $Ti_{1-x}Al_xN$ appears relatively unaffected by the orientation of the crack plane, as indicated by the overlapping error bars of square and triangular symbols in **Figure 7b**.

Although $Ti_{1-x}Al_xN(001)$ and $Ti_{1-x}Al_xN(111)$ cracked plate models return similar trends in mechanical properties, they also exhibit some remarkable differences in fracture mechanisms. Mode-I loading of TiN(111) reveals that (111) cracks are readily deflected along (001) planes (**Figure 8a**). The phenomenon is observed for all TiN(111) plate areas tested here. This suggests that the crack paths in TiN constantly follow the lowest-energy surface. In $Ti_{0.75}Al_{0.25}N(111)$ models, the crack wake exhibits a zig-zag pattern mainly formed of (111), (110), and (001) nanofacets (**Video #4**). The toughness and strength of $Ti_{0.75}Al_{0.25}N(111)$ are slightly improved compared to TiN(111), presumably due to formation of ligaments that slow down crack extension. A further increase in Al concentration progressively promotes nucleation and emission of dislocations from the crack tip of $Ti_{1-x}Al_xN(111)$. In $Ti_{0.4}Al_{0.6}N(111)$, the mechanism leads to formation of a stacking fault on the $(1\bar{1}1)$ plane, inclined 70.5º to the (111) crack surface, as indicated by the orange circle in **Figure 8b**. Slip-induced plastic deformation blunts the crack, thus relieving the high tip stress and enhancing the material resistance to fracture. However, an Al content greater than ~0.7 onsets plastic deformation at a "too" early stage of loading, which has a less positive effect on the mechanical properties of B1 $Ti_{1-x}Al_xN(111)$. This is reflected by a decline in the $K_{Ic}^{\infty}$ and $\sigma_f^{\infty}$ for $x>0.7$ (**Figure 7a,b**). Specifically, $Ti_{0.25}Al_{0.75}N(111)$ nucleates and emits the first dislocation at $K_I \approx$ 2.1 MPa √m. The event triggers a rapid sequence of dislocation glides (avalanche emission), effectively resulting in the formation of hexagonal (mainly B4-structured) domains (**Figure S12**).

Simulations done for crystal models with (001) and (111) cracks highlight that $Ti_{1-x}Al_xN$ alloys become more plastic with an increasing concentration of aluminum. In $Ti_{1-x}Al_xN(001)$, plastic deformation is primarily manifested through localized lattice distortions, amorphization, and phase transformation. Transformation at the tip of (001)-cracks in Al-rich $Ti_{1-x}Al_xN$ alloys is activated, and subsequently mainly driven, by tension (path illustrated in **Figure 1c**). However, the shear stress induced in the cubic $Ti_{1-x}Al_xN(001)$ lattice during mode-I loading contributes to the growth of the $B_k$ phase in Al-rich alloys (see example of $Ti_{0.15}Al_{0.85}N$ in **Figure S13a** in **SI**). Besides reducing the barrier ($\Delta E_{B1\rightarrow Bk}$) for tensile $B1\rightarrow B_k$ transformation (**Figure 3a**), an increased concentration of Al in $Ti_{1-x}Al_xN$ lowers the $\{110\}\langle 1\bar{1}0\rangle$ unstable stacking fault energy $\gamma_{usf}$ (**Table S2** in **SI**). This quantity is directly connected with the energy of activation of local $B1\rightarrow B_k$ transformation by shearing (**Figure S13b**). Although both $\Delta E_{B1\rightarrow Bk}$ and $\gamma_{usf}^{\{110\}<1-10>}$ decrease with increasing x, our simulations show that tensile-induced lattice transformation is the main plasticity mechanism in $Ti_{1-x}Al_xN(001)$ subjected to mode-I loading. Nevertheless, mixed mode-I/mode-II loading and/or presence of less sharp (001)-cracks could make dislocation-mediated plasticity as the dominant deformation mechanism (see results for TiN in Ref. [40]).

In $Ti_{1-x}Al_xN(111)$, the predominant mechanism for toughening and strengthening are the nucleation and emission of dislocations along $(1\bar{1}1)$ planes, resulting in the formation of stacking faults. We have indeed shown by MS calculations and experiments that the energy of formation of (111) stacking faults in $Ti_{1-x}Al_xN$ decreases with an increasing aluminum content: 1.43 $J\cdot m^{-2}$ for TiN, 1.0±0.2 $J\cdot m^{-2}$ for $Ti_{0.4}Al_{0.6}N$, and 0.7±0.3 $J\cdot m^{-2}$ for $Ti_{0.25}Al_{0.75}N$ [69]. Likewise, the *unstable* stacking fault energy ($\gamma_{usf}$) decreases monotonically with increasing x (**Table S3**). The reduction in stacking fault energies, leading to an increased ability of the alloys to slip on (111) planes, is the manifestation of a lower activation energy for cubic-to-hexagonal polymorphic transformation.

The different dislocation activity observed in $Ti_{1-x}Al_xN(001)$ and $Ti_{1-x}Al_xN(111)$ models can be explained with the help of Rice criterion [70]. According to Rice, the stress intensity factor $K_{Ie}$ indicates the stress condition necessary to emit dislocations from crack tips during mode-I loading. If $K_{Ie}$ is lower than $K_{Ic}$, the material is likely to activate slip-plasticity before fracture. Dislocation emission blunts the crack, which alleviates stress at the crack front and thus hinders crack propagation. We compute $K_{Ie}^{R}$ (the superscript "R" indicates Rice criterion for emission) accounting for the anisotropic elastic response of the crystal and the unstable stacking fault energy on the emission plane (equation 6b in Ref. [70]). The slip systems operative at the tip of (001) and (111) cracks are the $\{110\}\langle 1\bar{1}0\rangle$ and $\{111\}\langle 11\bar{2}\rangle$. The corresponding unstable stacking fault energies $\gamma_{usf}$ are given in **Tables S2 and S3**.

**Figure 9a** illustrates trends in $K_{Ie}^{R}$ calculated as a function of x in $Ti_{1-x}Al_xN(001)$ and $Ti_{1-x}Al_xN(111)$. Overall, the stress intensity $K_{Ie}^{R}$ decreases with increasing concentration of Al, suggesting that the likelihood of emission from crack tips increases with x. Neglecting the cases of $Ti_{0.15}Al_{0.85}N(001)$ and $Ti_{0.05}Al_{0.95}N(001)$ (that transform prematurely upon loading), a comparison of $K_{Ie}^{R}$ values shows that slip-plasticity is more favorable for (111) than for (001) cracks. The lowest values of $K_{Ie}^{R}$ (~2 MPa $\sqrt{m}$) are obtained for the $Ti_{0.5}Al_{0.5}N(111)$, $Ti_{0.4}Al_{0.6}N(111)$, and $Ti_{0.25}Al_{0.75}N(111)$ (green circles in **Figure 9a**). These systems also exhibit the highest $K_{Ic}^{\infty}/K_{Ie}^{R}$ ratio[4] between the fracture toughness $K_{Ic}^{\infty}$ and the stress intensity for emission $K_{Ie}^{R}$ (**Figure 9b**), which indicates high propensity to nucleate and emit dislocations. The results of K-controlled simulations confirm validity of the Rice criterion by showing the formation of stacking faults at the tip of (111) cracks in $Ti_{0.5}Al_{0.5}N$, $Ti_{0.4}Al_{0.6}N$, and $Ti_{0.25}Al_{0.75}N$. For $Ti_{0.25}Al_{0.75}N$ (**Figure S12**), the formation of the first stacking fault promotes subsequent slip events leading to a local B1→B4 lattice transformation. To summarize, ratios $K_{Ic}^{\infty}/K_{Ie}^{R} > 1.5$ are indicative of systems in which dislocation nucleation and emission is the primary plasticity mechanism. In all other defective $Ti_{1-x}Al_xN$ models (with $K_{Ic}^{\infty}/K_{Ie}^{R} \leq 1.5$), the simulations evidence brittleness or predominance of tension-induced lattice transformation.

[f4] Typically, the comparison is done between $K_{Ie}^{R}$ and $K_{Ic}^{G}$ calculated accoring to Griffith. However, the trends in $K_{Ic}^{G}$ vs x are qualitatively incorrect (see **Figure 2a** and **2b**).

### 3.4. Comparison with experimental fracture properties

Besides showing simulation results, **Figure 7** reports literature values of fracture toughness ($K_{Ic}^{exp}$) and fracture strength ($\sigma_f^{exp}$) measured for TiN and $Ti_{1-x}Al_xN$ coatings by microcantilever bending and micropillar splitting. The experimental values were collected for as-deposited $Ti_{1-x}Al_xN$ coatings with metal/N stoichiometry near unity ($K_{Ic}^{exp}$: [17-19, 30, 71-78] and $\sigma_f^{exp}$: [18, 19, 71-73]). Mechanical properties assessed by nanoindentation are omitted from the figure due to the bias introduced by residual film stresses [79] and mixed type of loading (compressive, tensile, shear) near indented regions [80].

The trends in fracture toughness and fracture strength determined by atomistic simulations are consistent with experiments (**Figure 7**). The properties measured for TiN are scattered over a relatively wide range: $K_{Ic}^{exp}$ ~1 – 3.5 MPa √m and $\sigma_f^{exp}$ ~1.5 – 5 GPa; effect that could be due to samples with different average grain sizes [81], different material density at grain boundaries, or different microstructural quality. Nevertheless, the results of simulations done on TiN models with (001) and (111) cracks ($K_{Ic}^{\infty}$=1.84±0.03 and 2.59±0.12 MPa √m; $\sigma_f^{\infty}$ = 2.37±0.13 and 2.54±0.18 GPa) are within the confidence ranges of the mean experimental values ($K_{Ic}^{exp}$=2.3±0.7 MPa √m; $\sigma_f^{exp}$=3.1±1.0 GPa). Consistent with the theoretical trends, the $K_{Ic}^{exp}$ and $\sigma_f^{exp}$ determined for $Ti_{1-x}Al_xN$ exhibit maxima for x around 0.6. Similarly, the machining performance and lifetime of $Ti_{1-x}Al_xN$-coated cutting tools increases monotonically with the concentration of Al up to x ~ 0.6–0.7 (see figure 7 in Ref. [45]).

In general, the fracture toughness and strength measured for polycrystalline samples cannot be compared directly to properties calculated for single-crystals. While our models elucidate transgranular fracture mechanisms, ex-situ (post-mortem) microscopy cannot reveal whether failure of $Ti_{1-x}Al_xN$ occurred intergranularly or transgranularly [74, 82]. On the other hand, the calculated $K_{Ic}$ and $\sigma_f$ trends are consistent with experiments (**Figure 7**). This suggests that atomic-scale plasticity and polymorphic competition largely control the fracture properties of $Ti_{1-x}Al_xN$ ceramics, irrespective of whether fracture is intergranular or transgranular. To support this argument, we present an analysis of the properties of grain boundaries in B1 $Ti_{1-x}Al_xN$ alloys.

The size, relative density, and structure of grain boundaries in B1-structured nitrides, carbides, and oxides vary largely depending on the synthesis parameters [83-90]. However, based on experimental information collected for B1 TMN [91-93], MgO [85, 94], and NiO [84], we can identify three relevant symmetric tilt grain boundaries: Σ3{112}[110], Σ5{210}[001], and Σ5{310}[001] (**Figure S14**). **Table S4 of the SI** shows that the energies ($E^{GB}$) calculated by MS for the Σ3 and Σ5 grain boundaries in B1 TiN (2.70 – 3.29 J/m$^2$) are in reasonable agreement with previous DFT values (1.73 – 2.13 J/m$^2$ [43, 95-98]). Both MS and DFT results return the relationship: $E^{GB}_{\Sigma5\{310\}} < E^{GB}_{\Sigma5\{210\}} < E^{GB}_{\Sigma3\{112\}}$. K-controlled simulations of grain boundary fracture are beyond the scope of this work. Nevertheless, trends in intergranular fracture toughness assessed using Griffith's criterion indicate that the resistance to fracture of the three boundaries decreases (exception made for Σ5{210}[001] in $Ti_{0.25}Al_{0.75}N$) with an increasing concentration of aluminum, as shown in **Table S5**. These results rule out the possibility that the enhancement in toughness and strength observed in experiments for $Ti_{1-x}Al_xN$ with x~0.6 arises from a bare improvement of cohesive forces between grains. Irrespective of whether cracks propagate along boundaries or across grains, the agreement between experimental and calculated properties in **Figure 7** indicates that aluminum-induced atomic-scale plasticity can positively affect both the intergranular and transgranular resistance to fracture of actual B1 $Ti_{1-x}Al_xN$ ceramics. On the other hand, $Ti_{1-x}Al_xN$ polycrystals with microstructures that differ greatly from typical PVD nitride coatings could exhibit qualitatively different trends in fracture properties.

To summarize, we highlight the positive contribution of atomic-scale plasticity to the fracture resistance of hard ceramics (H ≳ 20 GPa). The finding challenges the common belief that toughening mechanisms in brittle solids operate primarily at the microscale and that extrinsic toughening is the only route to improve the resistance to fracture of crystalline ceramics [67]. In this regard, recent experimental studies have shown that crystalline ceramics can be intrinsically toughened by artificially increasing the density of native dislocations [99-101]. Our work introduces an alternative strategy that exploits polymorphic competition to regulate phase-transformation toughening at the nano or sub-nano scale.

### 3.5. Polymorphic competition in alloys that spinodal decompose

The central hypothesis of this work – "polymorphic competition" as a strategy for designing ceramics with enhanced toughness – was inspired by the nanoscale stress-activated phase transformation observed in $ZrO_2$ and $ZrO_2$-based alloys. The occurrence of transformation plasticity in hard ceramics displaying polymorphic competition may have been previously overlooked, given the extremely small length scales involved (for example, ~Å-scale plasticity in $Ti_{0.4}Al_{0.6}N$, **Figure 4**). Our findings demonstrate that alloying a purely brittle ceramic (e.g., TiN) with a material that possesses various metastable phases (e.g., AlN) allows to tailor the extent of tension-activated transformation or shear-induced faulting. By compositionally modifying the energy barriers to selectively promote specific plasticity mechanisms, one could effectively enhance the intrinsic toughness of different families of ceramic materials.

Besides presenting polymorphic competition, B1 $Ti_{1-x}Al_xN$ solid solutions with intermediate content of Al undergo spinodal decomposition at elevated temperature [23]. The mechanism offers an additional degree of freedom to control mechanical properties. The properties of B1 AlN-rich / B1 TiN-rich semicoherent domains in spinodal decomposed $Ti_{1-x}Al_xN$ can be mimicked using TiN/AlN superlattice models. In a TiN/AlN superlattice structure, stress-activated phase transformation is facilitated (relatively to $Ti_{1-x}Al_xN$ solid solutions with equal composition) due to locally higher concentrations of AlN. This has been shown by recent tensile-strain simulations of defect-free TiN(001)/AlN(001) superlattices (Al/Ti ratio = 1 and bilayer periods $\Lambda$ = 1.25, 2.5, 5, and 10 nm) [62]. Stress-activated lattice transformation improves the plasticity of ceramic superlattices. However, analogous to our present observations for B1 $Ti_{1-x}Al_xN$ alloys, the beneficial impact of transformation plasticity on the theoretical strength and toughness of TiN/AlN superlattices can vary depending on the precocity and extent of the phenomenon. For example, simulations in Ref. [62] indicated that superlattices with $\Lambda > 2.5$ nm are overall less strong and less tough than structures with $\Lambda \leq$ 2.5 nm because of the complete AlN phase transitions activated at early stages of tensile strain. Moreover, the formation of incoherent hexagonal AlN / cubic TiN structures increases the interfacial strain, thus rapidly leading to crack nucleation and growth. In contrast, superlattices with an intermediate bilayer period ($\Lambda \sim 2.5$ nm) exhibit the best combinations of mechanical properties due to strain-mediated lattice distortions homogenously distributed throughout the material [62].

The knowledge gained by simulations on TiN/AlN superlattices [62] allows us to advance hypotheses on effects that spinodal decomposition could induce on the fracture toughness and fracture strength of (Ti,Al)N. The length-period of compositionally-modulated TiN/AlN domains increases with the annealing time and/or temperature. It is reasonable to assume that the fracture toughness of spinodal decomposed (Ti,Al)N alloys reaches a maximum for intermediate decomposition length periods, which would be similar to the behavior of TiN/AlN superlattices. The results of a recent experimental work corroborate our hypothesis by showing that $Ti_{0.4}Al_{0.6}N$ films annealed at 850 ºC are tougher than the as deposited samples (see figures 4a, 5, and 6 in [30]). An annealing temperature of 850 ºC onsets spinodal decomposition in $Ti_{0.4}Al_{0.6}N$ but is yet sufficiently low to prevent formation of the wurtzite hexagonal phase (figure 1 and figure 3c in [30]), which would be detrimental for hardness. Thus, the higher fracture toughness measured for annealed $Ti_{0.4}Al_{0.6}N$ samples by cantilever bending and verified by cube corner nanoindentation [30] is likely to arise from lattice transformation mechanisms initiated in AlN-rich domains. We conclude that the combination of polymorphic competition and spinodal instabilities could be exploited to simultaneously enhance the high-temperature hardness and fracture toughness of ceramic alloys.

## 4. Conclusions

Atomistic simulations elucidate fundamental mechanisms that govern the fracture toughness and fracture strength of $Ti_{1-x}Al_xN$ solid solutions. $Ti_{1-x}Al_xN$ alloys are taken as representative material systems to reveal and understand the effects of atomic-scale plasticity on the fracture properties of hard refractory ceramics. The results – validated by comparison with available experimental data of the mechanical properties of $Ti_{1-x}Al_xN$ – evidence that the energetic competition between polymorphs is a key parameter for controlling the onset and extent of phase transformation plasticity upon loading. Plasticity is shown to generally improve the alloys' resistance to fracture. However, an optimal combination of material strength and toughness is achieved for compositions that lead to moderate plastic deformation, as this allows accommodating a larger amount of mechanical stress prior to fracture. We envision that controlling polymorphic competition may become a valuable approach to enhance fracture resistance in various families of hard ceramics. Additionally, we argue that polymorphic competition in spinodal-unstable systems could be exploited to simultaneously enhance the high-temperature hardness and fracture toughness of ceramic alloys.

## Acknowledgements

The computations were enabled by resources provided by the National Academic Infrastructure for Supercomputing in Sweden (NAISS) at NSC in Linköping and PDC in Stockholm partially funded by the Swedish Research Council through Grant Agreements No. 2022-06725. Dr. Andrew Winters at the Department of Mathematics (LiU) and Dr. Luis Casillas Trujillo at NSC are gratefully acknowledged for technical support. We gratefully acknowledge financial support from the Swedish Research Council (VR) through Grants No. VR-2021–04426 and No. 2023-05358, the Competence Center Functional Nanoscale Materials (FunMat-II) (Vinnova Grant No. 2022–03071), the Swedish Government Strategic Research Area in Materials Science on Functional Materials at Linköping University (Faculty Grant SFO-Mat-LiU Grant No. 2009–00971), and the Knut and Alice Wallenberg Foundation through Wallenberg Scholar project (Grant No. 2018.0194). This work is supported by ERC Grant UNMASCC-HP, 101117758, funded by the European Union. Views and opinions expressed are however those of the author(s) only and do not necessarily reflect those of the European Union or the European Research Council Executive Agency. Neither the European Union nor the granting authority can be held responsible for them.

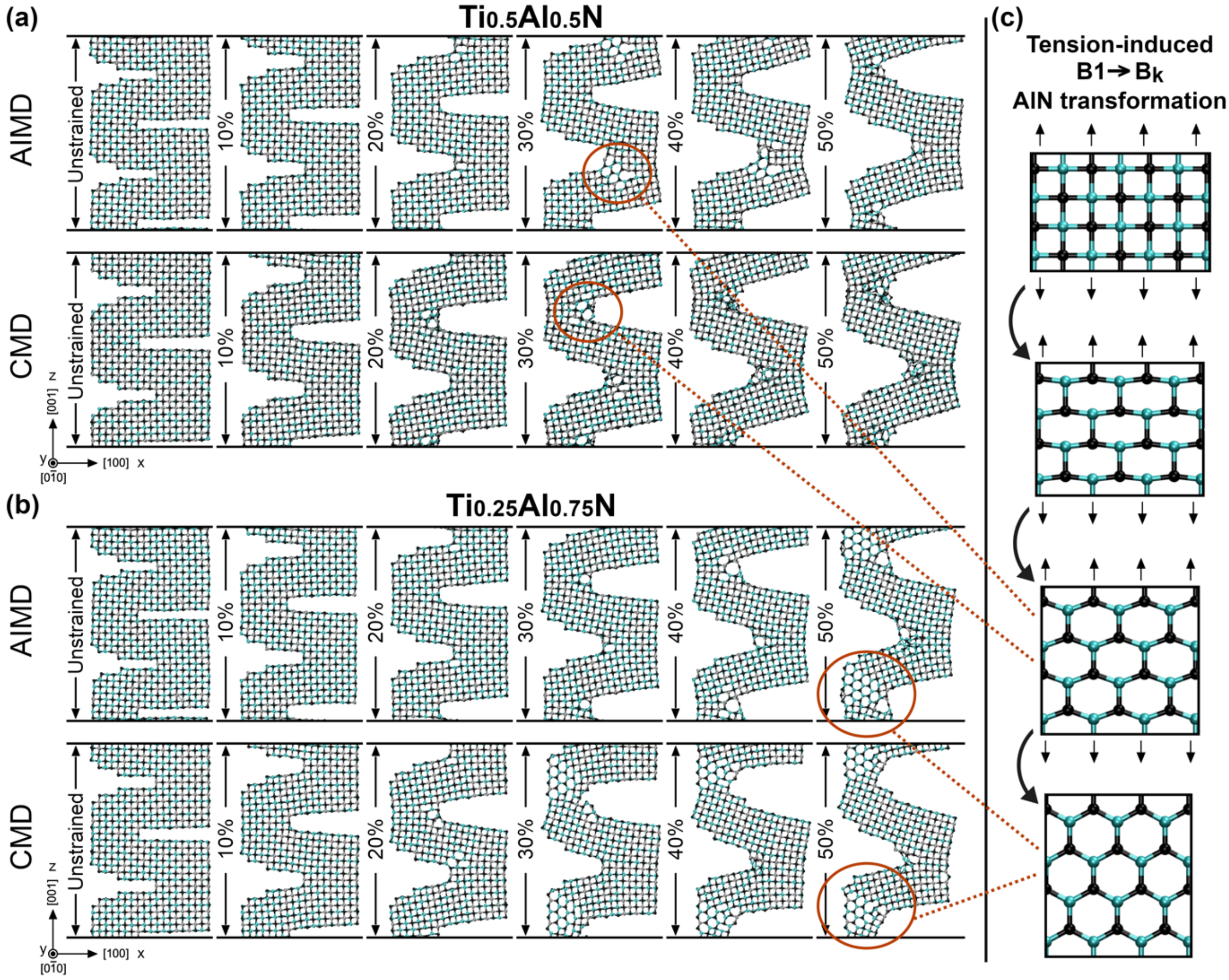

**Figure 1.** AIMD and CMD results of B1-structured bi-notched **(a)** $Ti_{0.5}Al_{0.5}N$ and **(b)** $Ti_{0.25}Al_{0.75}N$ supercell models subject to [001] elongation at 300 K. The structures are periodic along the y and z Cartesian directions. Vacuum regions separate supercell replicas along x. During AIMD simulations, self-healing of the notches is prevented by keeping eight helium atoms (not shown) at fixed position in the middle of the cavities. N, Al, and Ti atoms are colored in black, cyan, and silver, respectively. **(c)** Schematic representation of tension-induced B1→$B_k$ transformation in AlN. Orange circles and dashed lines indicate formation of $B_k$-like environments near the notches of vertically-elongated $Ti_{0.5}Al_{0.5}N$ and $Ti_{0.25}Al_{0.75}N$ supercells.

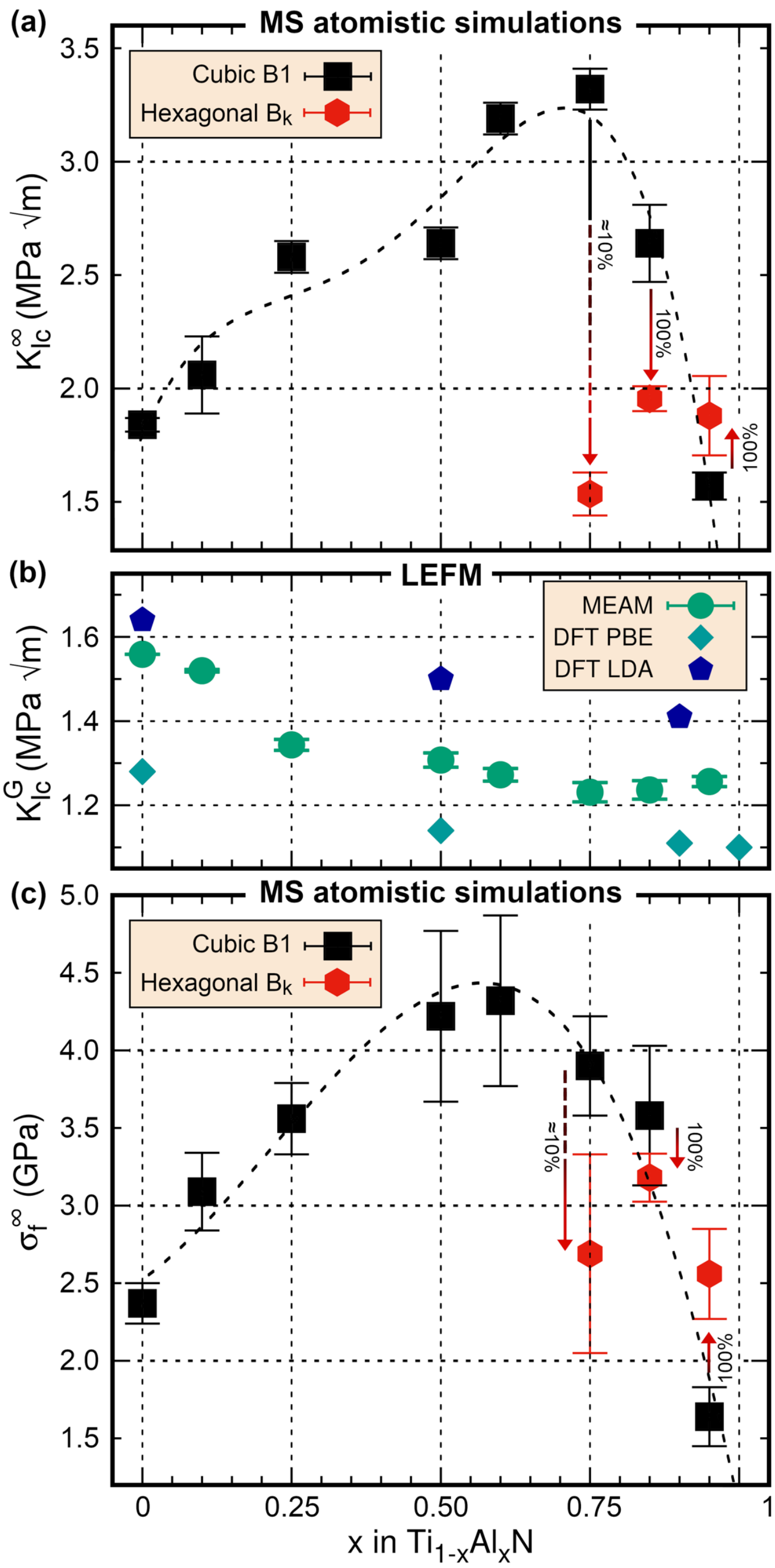

**Figure 2. (a)** Macroscale fracture toughness ($K_{Ic}^{\infty}$) and **(c)** macroscale fracture strength ($\sigma_f^{\infty}$) of $Ti_{1-x}Al_xN$ lattice models computed by MS $K_I$-controlled simulations as a function of the Al content x. Values of single-phase B1 $Ti_{1-x}Al_xN(001)$ are indicated by black squares. Red hexagons are mean values of fracture properties obtained for single-phase $B_k$ $Ti_{1-x}Al_xN$ with cracks on (1100) and (1120) planes. The vertical red arrows (labeled with percentages) indicate the probability of observing the B1→$B_k$ phase transformation during simulations of Al-rich (x > 0.7) alloys. The dashed curved black lines guide the eye across calculated trends. The extrapolated infinite-size values of $K_{Ic}^{\infty}$ and $\sigma_f^{\infty}$ are obtained by fitting $K_{IC}$ and $\sigma_f$ results collected for different plate areas (**see Section S1 in the SI**). For comparison with results in **(a)**, panel **(b)** shows the fracture toughness ($K_{Ic}^{G}$) values based on anisotropic linear elastic fracture mechanics (Griffith's theory). The alloys' unrelaxed surface energies $E^{(001)}$ and Stroh matrix elements $\Lambda_{22}^{-1}$ used for determination of Griffith's fracture toughness ($K_{Ic}^{G} = \sqrt{[2 \cdot E^{(001)} \cdot \Lambda_{22}^{-1}]}$) are obtained by both MS and DFT calculations (**see Section S2 in the SI**).

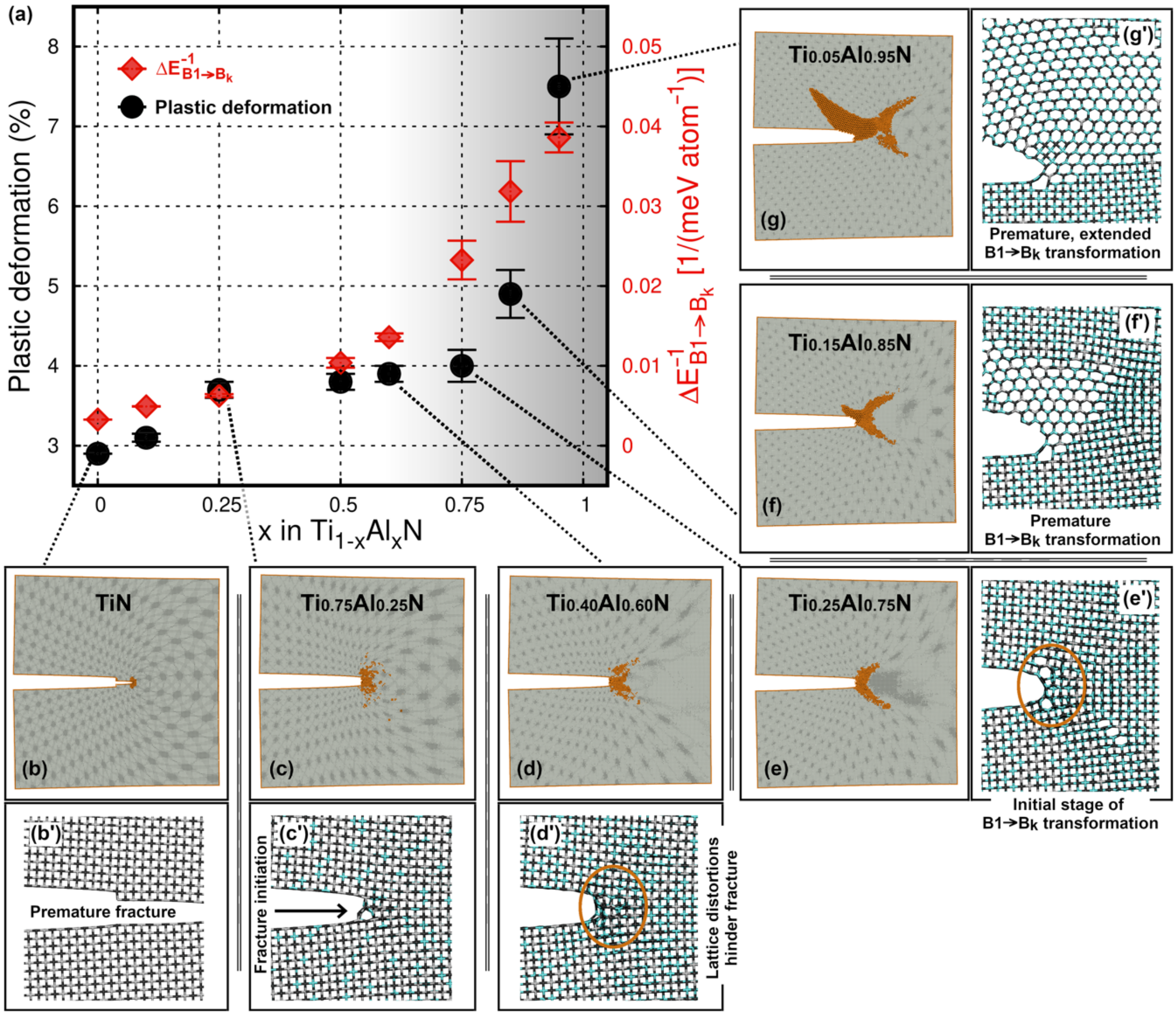

**Figure 3.** **(a)** Correlation between percentage of plastic deformation and inverse of the energy barrier $\Delta E_{B1\rightarrow Bk}$ for cubic-to-hexagonal $Ti_{1-x}Al_xN$ transformation separately calculated for 3D-periodic supercells. The energy landscape along B1→$B_k$ transformation paths are shown in **Figure S6 of the SI**. The percentage of transformed (non-cubic) lattice structure is determined using the common neighbor analysis on $Ti_{1-x}Al_xN$ plate models of area A≈1800 nm$^2$ (shown in **(b-g)**) at a stress-intensity factor $K_I$ = 2 MPa √m. In **(a)**, the white-gray color gradient at x≈0.6 indicates that both strength and toughness are initially positively affected by plasticity (white region), while they decrease for x > 0.7 (darker shade) due to premature stress-activated transformation. Atomistic models shown in **(b-g)** highlight relative portions of pristine B1-structured lattices (gray) as well as non-cubic plastically-deformed environments (orange). Panels **(b'-g')** are magnifications of regions around the crack tip, where N, Al, and Ti atoms are colored in black, cyan, and silver, respectively.

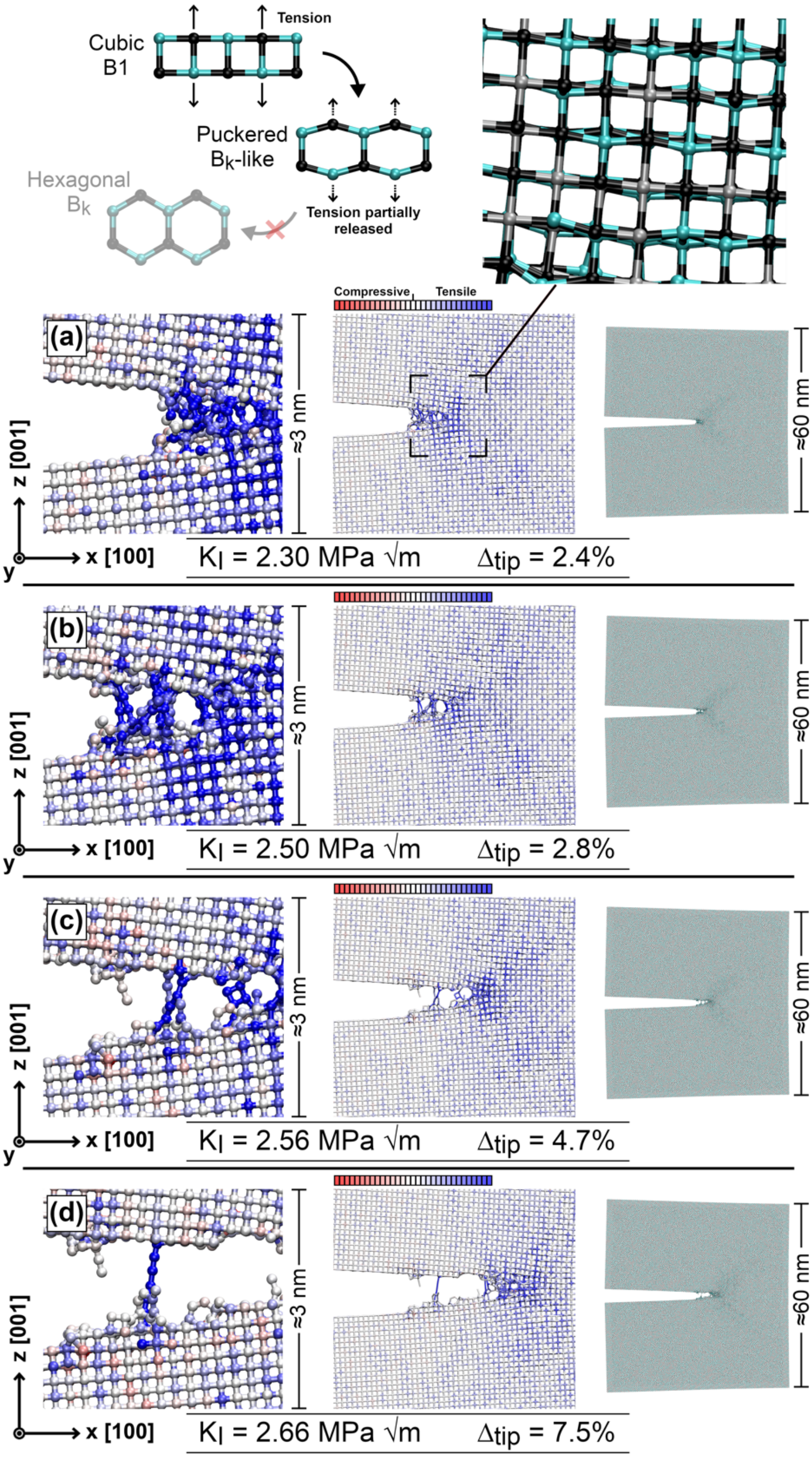

**Figure 4.** Al-induced strengthening and toughening mechanism in B1 $Ti_{1-x}Al_xN$. The figure shows as illustrative case a $Ti_{0.4}Al_{0.6}N$ cracked plate model with area ≈3600 nm$^2$. Note that atoms are colored according to the local tensile stress (see legend). **(a)** Formation of $B_k$-like domains around the crack tip at a stress-intensity factor $K_I$ = 2.30 MPa √m, which corresponds to a lateral advancement ($\Delta_{tip}$) of the tip of 2.4%. **(b,c,d)** Formation of ligaments with high load-carrying capacity. Acting behind the crack tip, the ligaments effectively toughen the alloy by obstructing crack propagation.

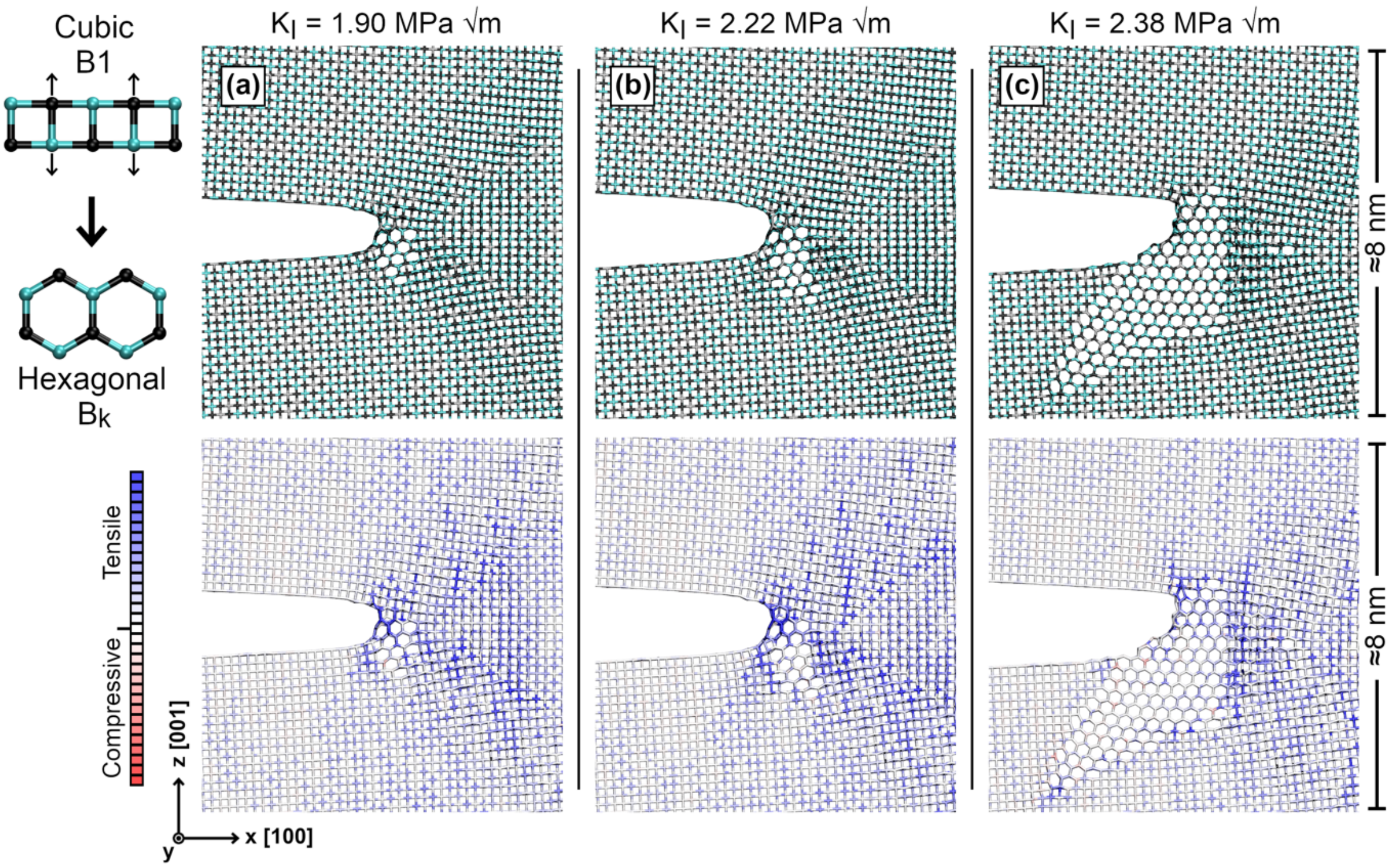

**Figure 5.** Simulation snapshots of B1-structured $Ti_{0.15}Al_{0.85}N(001)$ that locally transforms into $B_k$ during mode-I loading. The stress intensity factors indicated at the top are referenced to the B1 structure.

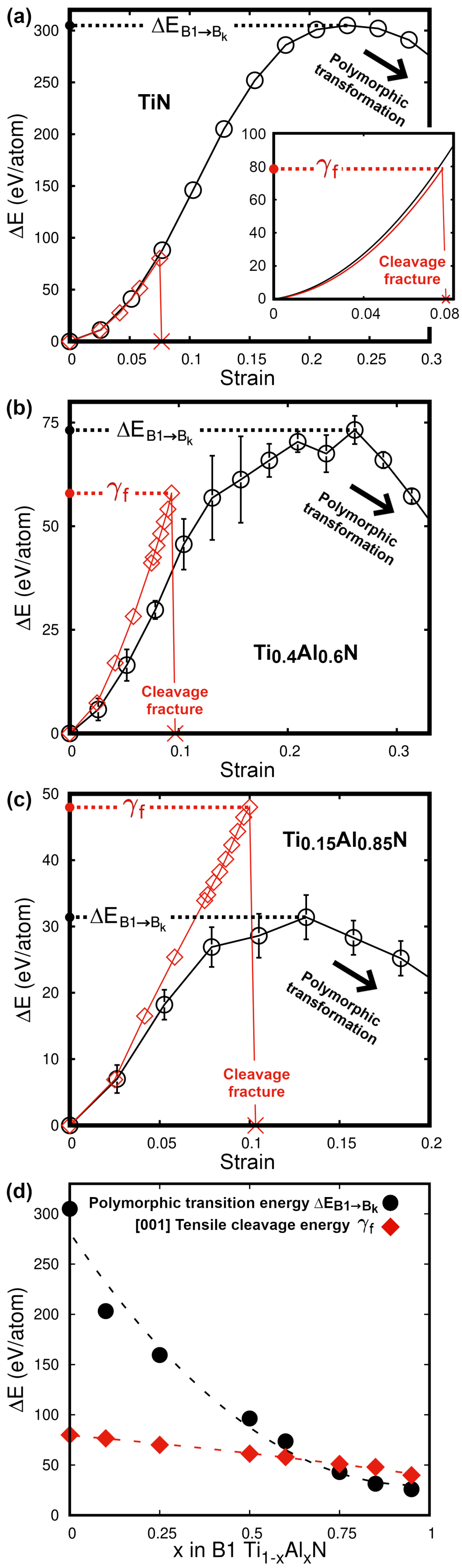

**Figure 6**. **(a-c)** Energetic competition between uniform tensile elongation along the [001] direction (red diamonds) and B1→$B_k$ polymorphic transformation energy (black circles) along the path shown in **Figure 1c**. Calculation details are given in **Section S3 of the SI**. The activation energies for B1→$B_k$ transformation and cleavage fracture of B1 $Ti_{1-x}Al_xN$ alloys on a {001} plane are summarized in **(d)**.

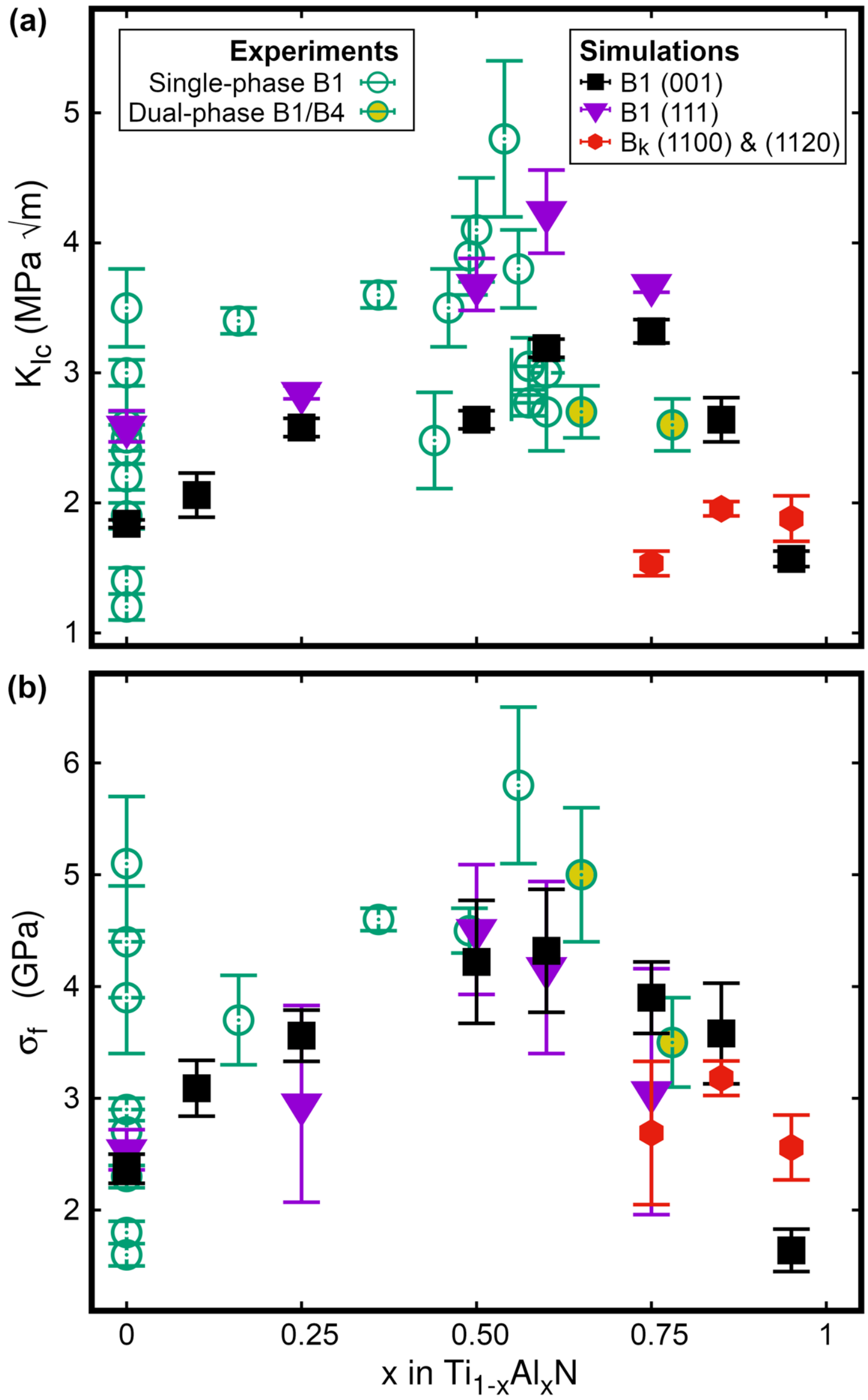

**Figure 7**. Comparison between fracture toughness $K_{Ic}^{\infty}$ and fracture strength $\sigma_f^{\infty}$ obtained by $K_I$-controlled simulations with fracture properties obtained by microcantilever bending and micropillar splitting of $Ti_{1-x}Al_xN$ as a function x. Note that experimental data colored in yellow were collected for dual-phase B1/B4 $Ti_{1-x}Al_xN$ [18]. The labels "B1 (001)", "B1 (111)", and "$B_k$ (1100) & (1120)" indicate simulation results obtained for B1-structured and $B_k$-structured $Ti_{1-x}Al_xN$ cracked plate models with atomically-sharp cracks on (001), (111), (1100) and (1120) planes. The results of simulations on B1 $Ti_{1-x}Al_xN$(001) and $B_k$ $Ti_{1-x}Al_xN$ are also shown in **Figure 2a,c**.

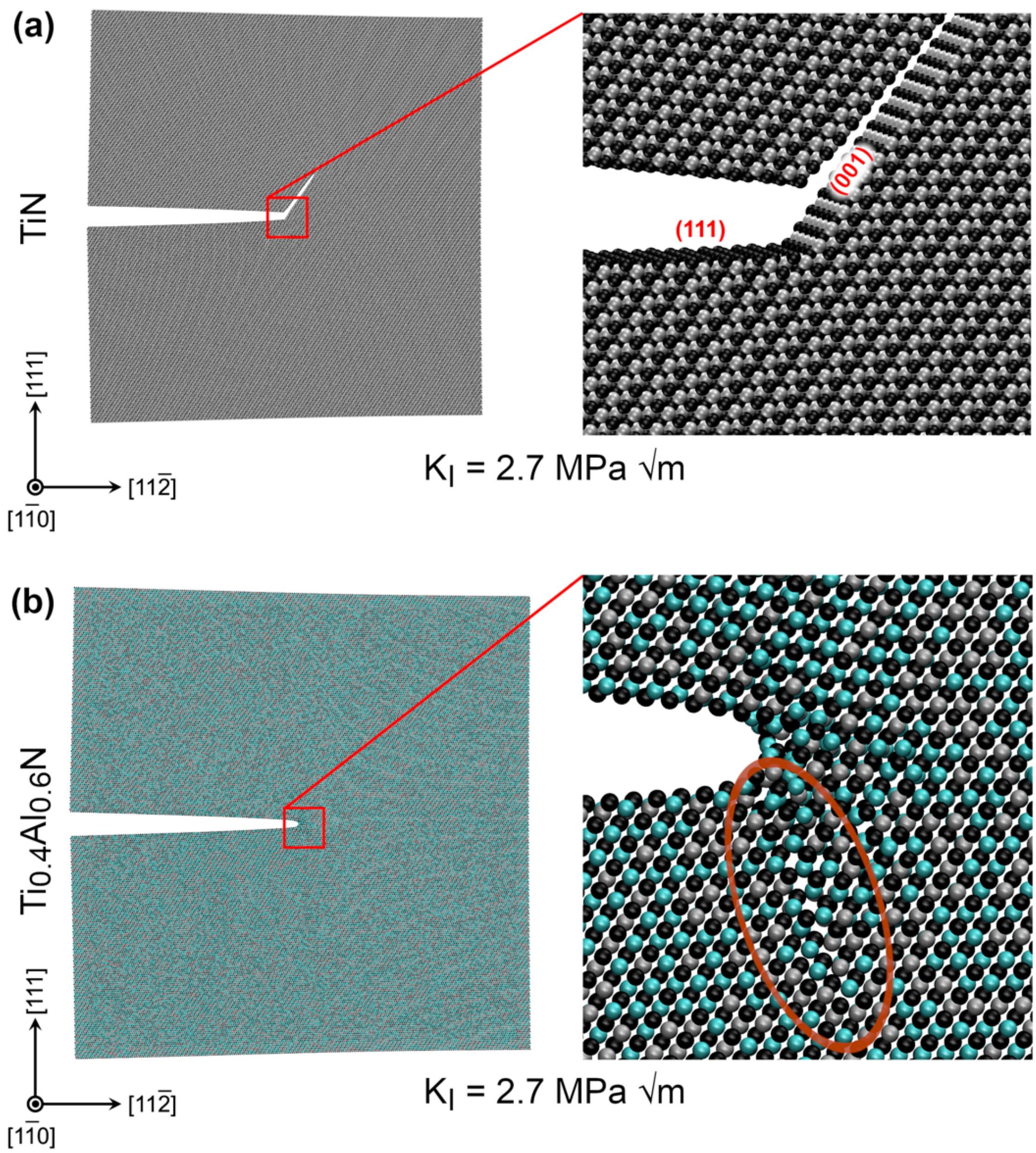

**Figure 8**. Simulation snapshots of $Ti_{1-x}Al_xN(111)$ cracked plate models subjected to mode-I loading. **(a)** An atomically-sharp crack on a {111} plane of TiN is readily deflected on a {001} plane. **(b)** Mode-I loading of $Ti_{0.4}Al_{0.6}N(111)$ induces nucleation and emission of a dislocation, indicated by the orange circle. N, Al, and Ti atoms are colored in black, cyan, and silver, respectively.

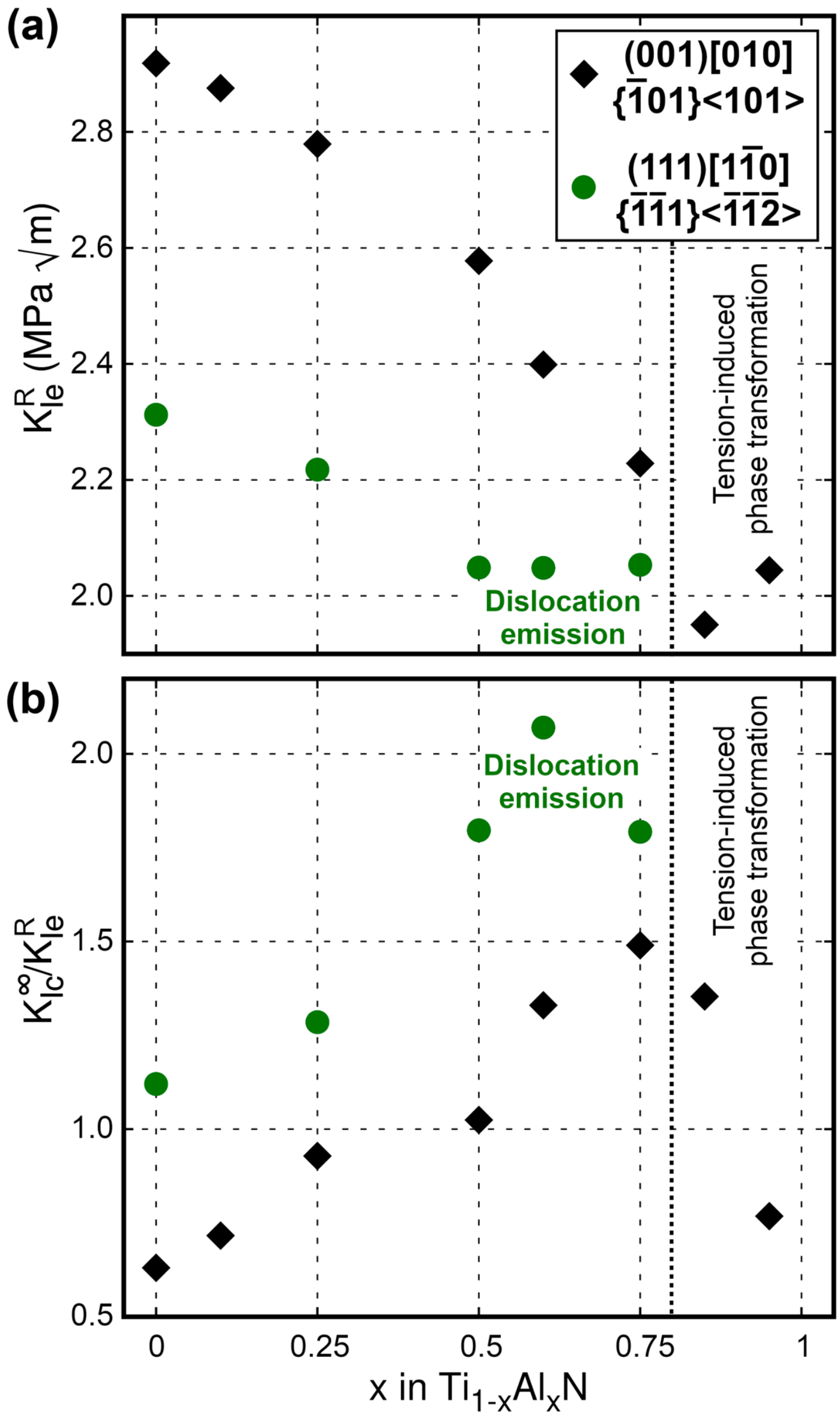

**Figure 9**. **(a)** Stress intensity factors for dislocation emission during mode-I loading ($K_{Ie}^R$) calculated according to Rice for $Ti_{1-x}Al_xN(001)$ and $Ti_{1-x}Al_xN(111)$ cracked plate models. In the legend, {plane}<direction> indicate the slip system for dislocation emission while (plane)[direction] denote the crack surface and crack line orientation. **(b)** Ratio of macroscale fracture toughness to Rice stress-intensity factor for emission during mode-I loading. The vertical dashed lines denote the compositional threshold above which the B1 structure transforms prematurely into $B_k$ (via tensile deformation path). The green labels "dislocation emission" indicate that slip-induced plasticity is observed in $Ti_{0.5}Al_{0.5}N(111)$, $Ti_{0.4}Al_{0.6}N(111)$, and $Ti_{0.25}Al_{0.75}N(111)$ models during K-controlled loading simulations. The formation of a stacking fault in $Ti_{0.25}Al_{0.75}N(111)$ is rapidly followed by slip events leading to a local B1→B4 lattice transformation (**Figure S12**).